# Abnormal Ionic Current Rectification Caused by Reversed Electroosmotic Flow under Viscosity Gradients across Thin Nanopores

Yinghua Qiu, [1,*,#] Zuzanna S. Siwy, [2] and Meni Wanunu[1,*]

[1] Department of Physics, Northeastern University, Boston, 02115, MA

[2] Department of Physics and Astronomy, University of California, Irvine, 92697 CA

yinghua.qiu@utah.edu, yinghua.qiu@hotmail.com

wanunu@neu.edu

# Present address: Department of Chemistry, University of Utah, Salt Lake City, 84112 UT

**Abstract:**

Single nanopores have attracted much scientific interest due to their versatile applications. The majority of experiments have been performed with nanopores being in contact with the same electrolyte on both sides of the membrane, while solution gradients across semi-permeable membranes are omnipresent in natural systems. In this manuscript, we studied ionic and fluidic movement through thin nanopores under viscosity gradients both experimentally and using simulations. Ionic current rectification was observed under these conditions, due to solutions with different conductivities filled across the pore under different biases caused by electroosmotic flow. We found that a pore filled with high viscosity solutions exhibited current increase with applied voltage in a steeper slope beyond a threshold voltage, which abnormally reduced the current rectification ratio. Through simulations, we found reversed electroosmotic flow that filled the pore with aqueous solutions of lower viscosities was responsible for this behavior. The reversed electroosmotic flow could be explained by slower depletion of coions than counterions along the pore. By increasing the surface charge density of pore surfaces, current rectification ratio could reach the value of the viscosity gradient across thin nanopores. Our findings shed light on fundamental aspects to be considered when performing experiments with viscosity gradients across nanopores and nanofluidic channels.

**Introduction**

Due to the confined space and surface charges, nanopores provide a versatile tool to tune ionic[1] and fluidic transport,[2] which is applicable to object sensing,[3-5] fluid pump designing,[6-7] energy conversion systems[8-10] and ionic transistors.[11-12] In a typical experimental setup, nanopores connect two conductivity cells filled with electrolytes, so that ionic, fluidic and analytes transport can only occur through the nanopores.

Even though the majority of reported nanopore experiments have been performed in symmetric electrolyte conditions, few measurements under gradients across the membranes have been reported as well, such as concentration gradients,[9-10, 13-15] pH gradients,[16-19] and viscosity gradients.[20-23] In nature, concentration gradients widely exist, including different salt concentrations across ion channels in cell membranes[24] or at the junction points between rivers and seas.[9] The investigations of concentration gradient could provide detailed physical information to various applications, like ionic gating[1, 15] and energy conversion.[8-10] For example, with an ion-selective membrane, energy conversion can be achieved through an entropy difference across the membrane. A pH gradient, as a kind of concentration gradient of proton or hydroxide ions, can affect the surface charge density along the pore.[25] Such gradients can be used to control fluid flow in the pore by modulating the surface charge density under different biases.[19] Due to the electroosmotic flow, the solution on the entrance side of the pore can be dragged into the pore.[2] The surface charge density can be controlled by the pH of the solution in the pore.[25]

Viscosity gradients appear usually with a kind of mass gradient across the pore, such as glycerol or dimethyl sulfoxide, which can control the viscosity and conductivity of the solution.[2] Viscosity gradients also exist widely in nature, like across cell membranes due to the crowded nature of inner cell environment [26-28] or in underground porous media, such as aquifers.[29] With the advantage of the high sensitivity of micro/nanopores, viscosity gradients across nanopores have been investigated before with potential applications in the detection of solution viscosity,[21] slowing down the speed of particles in resistive-pulse detection,[22, 30-31] as well as ionic transistors.[20-21, 32] Some groups have detected fluid flow under viscosity gradients with long micropores.[20-21, 32] Experiments show that the generated electroosmotic flow filled the pore with the solution from the entrance side, which can result in potential-dependent electrical resistance in the system i.e. current rectification in the current voltage (IV) curves. The current rectification caused by viscosity gradients is different from the traditional ionic current rectification, which is caused by enrichment or depletion of ions under different biases in pores with asymmetric geometries or asymmetric charge distributions.[11-12]

The effect of gradients across long pores can be understood by considering the two regions that differ in viscosity or ionic concentrations, without explicitly considering the liquid-liquid miscible region due to its small length scale. So, the current rectification, caused by the voltage-dependent electrical resistance of the system, has a ratio equal to the ratio of the solution conductivities.[20-21] However, when the pore length approaches nanoscale, the liquid-liquid miscible region cannot be ignored in the nanopore because it might occupy a large fraction of the pore volume, a condition that has been seldom considered in the literature.[20]

In this work, experiments and simulations of ionic and fluidic transport through silicon nitride (SiN) nanopores as thin as 50 nm had been conducted under viscosity gradients established between water and glycerol solutions. In order to minimize the influence of the pore shape on ionic transport, lower ratios of the Debye layer thickness over the pore diameter were considered through combination of nanopores with ~20 nm in diameter and electrolyte solutions with higher concentrations than 0.5 M.[2] Under viscosity gradients, abnormal current rectification could be found. With voltage increasing, the nanopore rectified more obviously until a threshold, which started the onset of a steeper current increase with voltage and reducing the current rectification ratio abnormally. From numerical simulations, we found that the voltage-dependent ionic current rectification was resulted from the reversed electroosmotic flow caused by excess of coions along the pore axis. This finding is closely related to the ionic and fluidic movement in the pore under viscosity gradients, which has potential applications in the design of ionic circuit and electroosmotic flow pumps, as well as detection of solution viscosities. When abnormal current rectification occurs, the fluid near the pore walls and in the center of the pore move in opposite directions, which could lead to mixing of the fluids from both sides of the pore in the confined space. In this case, the nanopore could be used as a nano-reactor to investigate dynamic interactions between the analytes from both sides of the pore, and probe chemical reactions and products.

**Experimental and Simulation Methods**

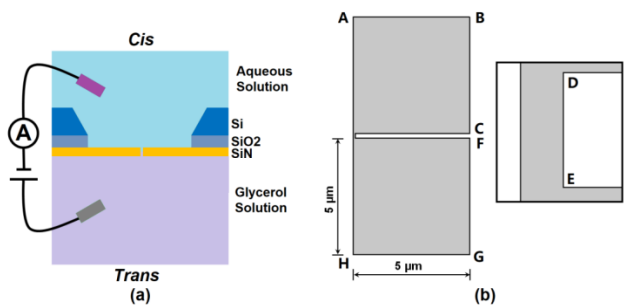

Figure 1. (a) Scheme of the experimental setup with a nanopore placed between two chambers of a conductive cell. (b) Scheme of the system used in simulations. Zoomed-in pore region is shown on the right of panel (b). The ground electrode was placed in the cis chamber of (a) and on AB boundary of (b).

**Experimental method.** The scheme of the experimental setup is shown in Figure 1a. Single nanopores with ~20 nm in diameter were drilled using a transmission electron microscope (JEOL 2010F) in 50-nm-thick freestanding SiN membranes. The SiN membranes (20~50 μm square windows) were fabricated on a 5×5 mm$^2$ silicon chip using a previously described process[33]. Briefly, low-pressure chemical vapor deposition was used to deposit a SiN layer on 500 μm thick silicon wafer with <100> orientation. Prior to the SiN deposition, in order to reduce the electrical capacitance noise, a SiO$_2$ barrier layer with 2.5 μm in thickness was formed through thermal oxidization. Before ionic current measurements, all drilled nanopores were cleaned with hot piranha solution (1:1 H$_2$SO$_4$/H$_2$O$_2$) for 15 minutes, followed by hot deionized water for 10 minutes, and drying under vacuum. A custom polytetrafluoroethylene (PTFE) conductivity cell was used to accommodate the SiN chips with nanopores. Quick-curing silicone elastomer was used to seal the gap between the chip and the cell, as well as to reduce the capacitive noise. Solutions were prepared using 99.0% KCl (Fisher Scientific, IN) in deionized (DI) water (purified using a Millipore system), as well as mixed deionized water and glycerol solutions (Amresco, Solon, OH). Solution conductivities (Table S1) were measured with a Fisher Scientific Accumet Basic AB30 conductivity meter at room temperature. Current–voltage curves were acquired with an Axopatch 200B amplifier and Digidata 1200 (Molecular Devices, Inc.), and recorded using a custom LabVIEW software. The recorded data were digitized and recorded at a sampling frequency of 250 kHz after applying a low-pass Bessel filter of 10 kHz. Two homemade Ag/AgCl electrodes were used to apply voltages and measure currents. The ground electrode was put in the chamber filled with aqueous solution, and the working electrode was placed on the other side of the pore in contact with the water/glycerol mixture. Voltage values were changed between –3 V and +3 V with a 0.1 V step. 0.5, 0.75 and 1.0 M KCl solutions pH 10 adjusted with 10 mM Tris base were used. KCl solutions in water/glycerol mixtures were prepared with 20%, 40% and 60 % glycerol in weight ratio to consider the cases with high viscosities. During the experiments, the chambers were cleaned with DI water twice after each detection, and rinsed with the solution to be used twice before next detection. For each case, detection was conducted 5 minutes after the solution was filled, and the current voltage curve was obtained by averaging at least 6 runs. Au nanoparticles (10 nm in core size) with polyethylene glycol (PEG)$_{5000}$ modification were purchased from Sigma-Aldrich (Sigma–Aldrich, St. Louis, MO). Particle size was detected as 33.8 nm in diameter with NanoBrook 90Plus PALS Particle Size Analyzer (Brookhaven Instruments Corp., Holtsville, NY). In the detection, 1 μL of as obtained particle solution was added to 70 μL KCl solution.

**Simulation method.** Simulations of ionic transport and fluid flow under viscosity gradients were conducted through solving coupled Poisson-Nernst-Planck (PNP) and Navier-Stokes (NS) equations with COMSOL Multiphysics 5.2. The scheme of the simulation system is shown in Figure 1b. All boundary conditions were listed in Table S2.[32, 34] In this work, the length of the cylindrical pore was varied from 20 to 1000 nm, and its diameter ranged between 5 and 200 nm. The surface charge density of the pores was chosen as –0.01 C/m$^2$ to simulate the case of SiN membranes in pH 10 solutions.[35-38] For the inner pore surface, 0.1 nm mesh size was used to take into account the effect of electrical double layers and assure the convergence of systems.[25] The mesh of 0.5 nm was chosen for the charged boundaries of the reservoirs to lower the memory cost during simulations. KCl, KF, LiCl and NaCl were considered in the simulations. Diffusion coefficients of the ions in water used in this work were assumed equal to the bulk values as show in Table S3.[39] The salt concentration was set from 0.5 to 1 M. Voltages applied in each case were from –3 to 3 V with a 0.2 V step.

In the system, viscosity gradients were considered through adding glycerol molecules in the system. The dielectric constants of pure water and pure glycerol were set as 80 and 42.5,[40] respectively. Glycerol molecules were simulated as neutral particles and put in one reservoir. The physical properties of the mixed solutions were calculated with the mass ratio of the glycerol[32] according to the equations given below. The density,[41] dielectric constant of the solution[42] and viscosity[41] were calculated using equation 1, 2 and 3, respectively. The dependence of diffusion coefficient of the glycerol molecule on its concentration was considered with equation 5.[43] The diffusion coefficients of ions in mixed solutions were calculated by equation 6.[44]

$$\rho_{solution} = x\rho_{glycerol} + (1-x)\rho_{water} \tag{1}$$

$\rho_{solution}$, $\rho_{glycerol}$ and $\rho_{water}$ are the density for a mixed solution of glycerol and water, pure glycerol and pure water, respectively. $x$ is the weight ratio of glycerol in the mixed solution.

$$\varepsilon_{solution} = x\varepsilon_{glycerol} + (1-x)\varepsilon_{water} \tag{2}$$

$\varepsilon_{solution}$, $\varepsilon_{glycerol}$ and $\varepsilon_{water}$ are the dielectric constant for mixed solution of glycerol and water, pure glycerol and pure water, respectively.

$$\mu_{solution} = \mu_{water} \cdot \left[\frac{1-x/C_m}{1-(k_0 C_m - 1)\cdot x/C_m}\right]^{-2.5C_m/2 - k_0 C_m} \tag{3}$$

$\mu_{solution}$ and $\mu_{water}$ are the viscosity for mixed solution of glycerol and water, as well as pure water, respectively. $C_m$ acts as the volume fraction for the dispersed particles, like glycerol molecules, at which the solution viscosity reaches an infinite value. $k_o$ accounts for the repulsive colloidal forces among dispersed particles which is related to the hydrodynamic forces.[41]

$$k_0 = -0.012T + 4.74 \tag{4}$$

$T$ is absolute temperature, $C_m = 0.74$ and $C_m = 1.2$ were used for the solutions with weight ratio of glycerol less than 60% and equal to 60%.[41] The obtained viscosity values with different glycerol percentages are shown in Figure S1 with comparison with experimental data.[45]

The diffusion coefficient of glycerol molecules in mixed solutions was described by equation 5, which was a linear fitting of the experimental results.[43]

$$D = (9.986 - 9.802x) \cdot 10^{-10} \left[ m^2/s \right] \tag{5}$$

Based on Stokes-Einstein equation[44] $D = \dfrac{\overline{R}T}{N_A} \dfrac{1}{6\pi\mu a}$, in which $\overline{R}$ is the gas constant, $N_A$ is Avogadro's number, $a$ is the radius of moving particle. The diffusion coefficient of ions in mixed solution can be calculated as following:

$$D_{i-solution} = \dfrac{\mu_{water}}{\mu_{solution}} D_{i-water} \tag{6}$$

$D_{i-solution}$ and $D_{i-water}$ are the diffusion coefficients of ions in mixed solution and water, respectively.

## Results and Discussions

Behaviors of large-aspect-ratio pores in contact with viscosity gradients were recently probed experimentally and numerically with single mesopores in polymer films.[21] Electroosmotic flow (EOF) caused by the negatively charged pore walls could drag the solution of the reservoir on the entrance side into the mesopores to induce the voltage-dependent electrical resistance. These pores exhibited ionic current rectification equal to the ratio of the two solutions' conductivities. This electroosmotically driven ionic current rectification could be fully predicted and easily tuned by changing the conductivity of one solution and holding that for the other solution. The experiments showed that for long pores, the liquid-liquid miscible region between water and 40% glycerol solution could be ignored due to its minute length scale. Herein we considered the simplified cases with pores under different viscosity gradients without the consideration of the liquid-liquid miscible regions. Hence, theoretical prediction for ionic current rectification could be given via equations of access resistance on one side ($R_{ac}$)[46] and pore resistance ($R_p$):[36]

$$R_{ac} = \dfrac{1}{2\kappa D} \tag{7}$$

$$R_p = \dfrac{4L}{\pi\kappa D^2} \tag{8}$$

where $\kappa$ is conductivity of the solution, $D$ is diameter of the pore and $L$ is length of the pore.

Pores were negatively charged so that the direction of electroosmotic flow was expected to be determined by the direction of cations' migration. In the electrode configuration chosen, the working electrode was placed in the medium of low conductivity as shown in Figure 1. Consequently positive voltages were expected to fill the pore with the low conductivity medium, and $|I_-/I_+|$ was expected to be larger than 1. The ionic current rectification (*ICR*) ratio could be calculated as:[11]

$$ICR = \left|\frac{I_-}{I_+}\right| = \left|\frac{V/R_-}{V/R_+}\right| = \frac{R_+}{R_-} = \frac{R_{ac-water} + R_{p-solution} + R_{ac-solution}}{R_{ac-water} + R_{p-water} + R_{ac-solution}} \quad (9)$$

$$\frac{\kappa_{water}}{\kappa_{solution}} = \frac{D_{i-water}}{D_{i-solution}} = \frac{\mu_{solution}}{\mu_{water}} \quad (10)$$

in which $I_+$ and $I_-$, $R_+$ and $R_-$ are the current and resistance under positive and negative voltages, respectively. $V$ is the applied voltage. $R_{ac-water}$ and $R_{ac-solution}$, stand for the access resistance on the side in contact with aqueous and mixed solution, respectively. $R_{p-water}$ and $R_{p-solution}$ are the pore resistance assuming the pore is filled with water and mixed solution. $\kappa_{water}$ and $\kappa_{solution}$ are the conductivity of the aqueous and mixed solutions, respectively. The viscosity parameters were taken from the literature.[45]

Figure 2 shows the predicted ICR ratios under viscosity gradients between aqueous solution on one side and mixed glycerol solutions with 20%, 40% or 60% in weight ratio of glycerol on the other side of the pore. When the pore length is short ($L/D<0.1$), access resistance on both sides of the pore dominates the total resistance. In this case, equation 9 predicts ICR≈1 i.e. no current rectification appears. This case had been experimentally found in ultra-thin MoS$_2$ pores subjected to a viscosity gradient.[22] With the pore length increasing ($0.1<L/D<10$), the pore resistance becomes more significant, leading to ICR>1 which means more obvious current rectification. Finally, when the pore length is ~10 times of its diameter, the ICR ratio reaches its maximum value equal to the viscosity ratio across the pore, as shown before.[20-21] However, when the length of the pore belongs to an intermediate range, such as less than a hundred of nanometers with a diameter as 20 nm, the liquid-liquid miscible region may extend across the pore due to diffusion of solvent molecules. In this case, the solution property is not uniform, that means the conductivity of the solution in the pore can be higher than that of glycerol solution but lower than that of aqueous solution. From the literature, there is little consideration of the cases of pores with a medium length under viscosity gradients.[20] We will study that due to its sensitivity to applied voltages, new transport phenomena can be observed.

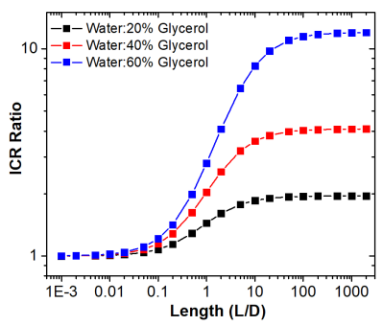

Figure 2. Ionic current rectification (ICR) ratios as a function of the nanopore aspect ratio predicted through equations 7-10 under three viscosity gradients.

In this work, ionic current behaviors through nanopores under viscosity gradients had been investigated through experiments and COMSOL simulations. For the case with a viscosity gradient between aqueous and 40% glycerol solutions, Figure S2 shows the distribution of glycerol molecules in the pore and the corresponding viscosity distributions from simulations. When a positive voltage was applied as 0.4 V, the EOF dragged the 40% glycerol solution to the pore, which made the concentration of glycerol molecules higher than the cases without a voltage and with a negative voltage. Therefore, the solution was more viscous. Because the pore was very short, there is no plateau regions in the concentration distribution of glycerol, which changed monotonously and was different from that in long pores.[16]

Experimental investigation of ionic current behaviors under viscosity gradients had been done with SiN pores with ~20 nm in diameter and 50 nm in length. As shown in Figure 3a, in the cases without viscosity gradients, the current depended linearly on the applied voltages. When the viscosity gradient appeared between aqueous and 40% glycerol solutions, current rectification happened correspondingly.[20] The ICR ratio is plotted in Figure S3. A peak ratio was found as ~1.9 in 1 M KCl solutions, which was lower than the conductivity ratio (Table S1) and viscosity gradient, suggesting the pore was filled with a mixture of water and glycerol.[20] For negative voltages, the EOF filled the pore with aqueous solution which resulted in a current similar to that of the aqueous case. Under positive biases, EOF dragged the glycerol solution into the pore, which induced a smaller current. Under lower voltages (<1.25 V), the current rectification ratio increased with voltage.[11] However, an interesting phenomenon appeared when the voltage reached ~1.25 V. A second steeper current slope appeared which suppressed the current rectification ratio until 1. Both the increased current and the decreased ICR ratio were unexpected, because usually larger EOF$^2$ under a stronger electric field can fill the pore with more viscous solutions.[20-21] Here, we call it abnormal current rectification. In order to roughly evaluate the location of the appearance of the second slope, i.e. the turning point, ICR ratio was plotted with voltage. The phenomenon of abnormal current rectification could be easily repeated with different SiN pores, as shown in Figure S4. Please note that due to the difference in pore geometry and opening diameter, the location of the turning point for different pores may be different slightly.

Different electrolyte concentrations and viscosity gradients were considered in the experiment. The abnormal current rectification happened with KCl concentration from 0.5 M to 1.0 M (Figure S5 and S6). Results in Figure 3b and S3a indicate that a lower salt concentration could provide a larger ICR ratio. This was due to the stronger electroosmotic flow in the pore[47] which was related with a thicker Debye layer and a higher surface potential.[25] The turning point in a more diluted solution was located at a little higher voltage among the comparison of ICR ratio shown in Figure 3b. For the cases with different viscosity gradients across the pore, higher viscosity gradients produced more obvious current rectification because of the larger difference in conductivities between viscous and aqueous solutions. From the ICR ratio plot, the turning point depended on the weight ratio of the glycerol closely. (Figure S3b and S6)

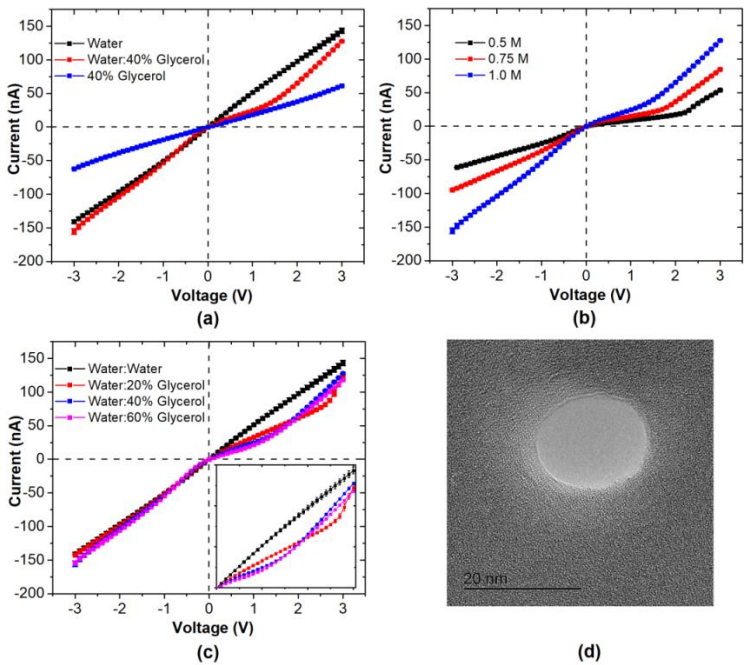

Figure 3. Experimental current voltage curves through a single SiN pore with ~18 nm in diameter and 50 nm in thickness. (a) Recording for cases of different viscosity gradients across the pore with 1 M KCl pH 10. (b) IV curves from cases with KCl in different concentrations at pH 10 under the same viscosity gradient between aqueous and 40% glycerol solutions. (c) IV curves from cases of 1 M KCl pH 10 solutions under different viscosity gradients. (d) TEM image of the pore used in the experiments.

In order to understand the mechanism of the abnormal current rectification, COMSOL simulations were conducted with a consideration of viscosity gradients across the nanopore with 20 nm in diameter, 50 nm in thickness, and –0.01C/m$^2$ in surface charge density. The same electrolyte concentrations and viscosity gradients as studied experimentally in Figure 3 were considered. Current voltage curves obtained from simulations are shown in Figure 4. The ICR ratios under different KCl concentrations and viscosity gradients are plotted in Figure S7. The simulation results were in excellent agreement with the experiment. Please note that the absolute values of the current rectification ratios in the simulations were a little less than those obtained in the experiments. This may be due to the small difference of pore shape or surface charge density used in the simulations from the real cases.[33, 35, 48]

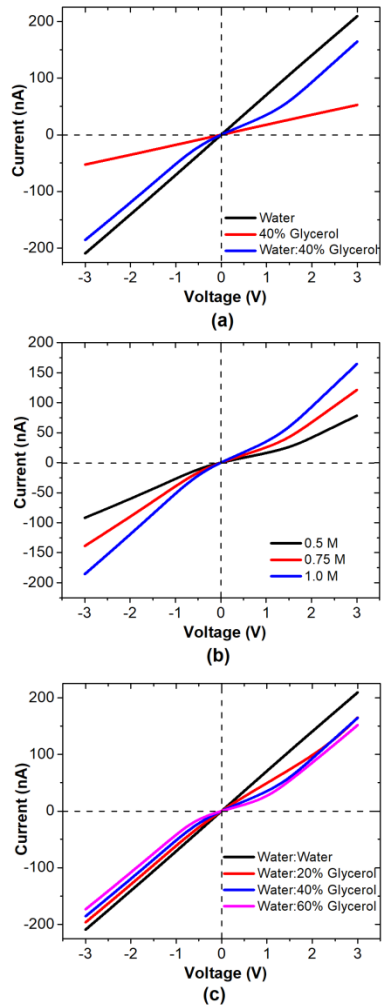

Figure 4. Simulation data of current voltage curves through a pore with 20 nm in diameter and 50 nm in thickness. (a) IV curves from cases of different viscosity gradients across the pore with 1 M KCl. (b) IV curves from cases with KCl in different concentrations under the same viscosity gradient between aqueous and 40% glycerol solutions. (c) IV curves from cases of 1 M KCl solutions with different viscosity gradients. The boundary conditions are listed in Table S2.

As the next step, the current voltage curves with detailed contributions of cations and anions were analyzed under the viscosity gradient with aqueous and 40% glycerol solutions, as shown in Figure 5a and 5b. The contributions of $K^+$ ions under no viscosity gradients shared almost the same value at different voltages, i.e. ~51%, a little larger than 50% due to the enhanced concentration caused by negatively charged surfaces.[25] While, in the presence of a viscosity gradient, the contribution of $K^+$ ions monotonously decreased from –1 V to 3 V to a level significantly below 50%, and remained at a relative constant value of 52.5% at voltages below –1 V. Similar results had also been obtained with different KCl concentrations and viscosity gradients. For the cases with different KCl concentrations, the percentage distributions of $K^+$ ions had the same trend. While, for the cases with different viscosity gradients, a higher viscosity gradient had a larger decrease in the $K^+$ ion contribution to the total current.

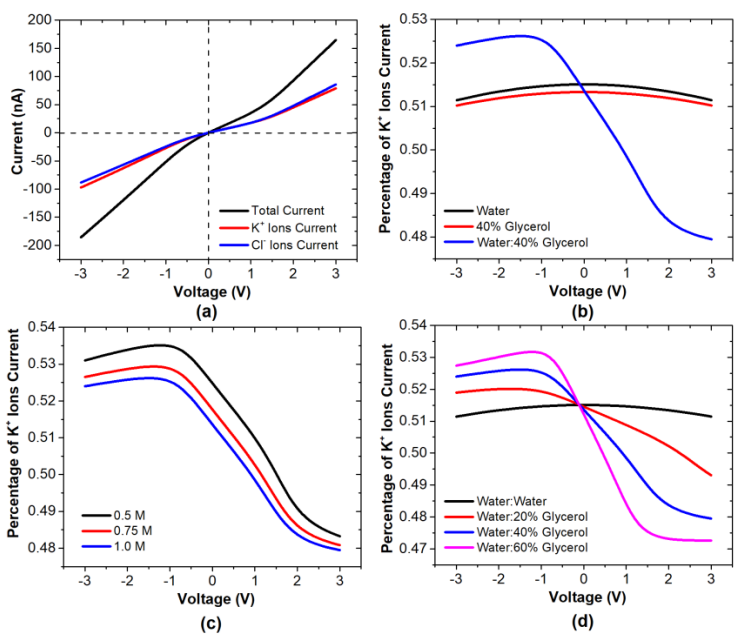

Figure 5. Simulation data of ionic current through a nanopore. (a) IV curves with detailed contributions from cations and anions in 1 M KCl solution under the viscosity gradient between aqueous and 40% glycerol solutions. (b) Contributions of $K^+$ ions to the total current in 1 M KCl solution in aqueous and glycerol solutions, as well as under the viscosity gradient. (c) Contributions of $K^+$ ions to the total current in KCl solutions of different concentrations under the viscosity gradient between aqueous and 40% glycerol solutions. (d) Contributions of $K^+$ ions to total current in 1 M KCl solution under different viscosity gradients.

The decreased contribution of cations to the total ion current under viscosity gradients was surprising, because SiN nanopores were expected to be negatively charged. Counterions were attracted to the pore surface to form electric double layers[25] which made $K^+$ ions the dominated current carriers, i.e. 51% current contribution in the aqueous and the 40% glycerol cases without gradients. Different electrolytes had been used to consider the influence of the ionic mobility. As shown in Figure S8, ionic mobility didn't have obvious effect on the abnormal current rectification phenomenon. In order to explain why the contribution of $K^+$ ions to the total current decreased at high positive biases, we investigated the ionic concentration distributions and electroosmotic flow in the pore which could influence the ionic current directly.[11, 49-50] As shown in Figure 6a and 6b, concentration of both ions was strongly voltage-dependent, and at high voltages concentration polarization became obvious. With the increase of applied positive voltage, the concentration of $Cl^-$ ions became higher than that of $K^+$ ions. The increase of co-ions concentration was indeed induced by the viscosity gradient, because in symmetric aqueous conditions, counterions shared almost the same concentration as coions at all voltages (Figure S9c and 9d). We thought this was caused by the different ionic mobility change along the pore axis for $K^+$ and $Cl^-$ ions. Under positive voltages, $K^+$ ions moved from solution with a higher viscosity to aqueous solution. Their ionic mobility increased along the axis. While, $Cl^-$ ions passed through the pore in the opposite direction, with their mobility decreasing monotonously during the translocation process. Consequently, $Cl^-$ ions could have a larger concentration than $K^+$ ions due to the slower depletion at the end of the pore connected with a high viscosity solution, as shown in Figure 6a and Figure S9a. We also found that comparing with the case of symmetric solutions, higher depletion of $K^+$ ions happened at the pore entrance under viscosity gradients due to the increase of its mobility along the pore axis (Figure S10).

However, $K^+$ and $Cl^-$ ions shared almost the same mobility at the same locations in the pore. A small difference in their concentration cannot produce such an obvious current increase. The voltage-dependent relative concentrations of cations and anions prompted us to look into details of magnitude and direction of electroosmotic flow in the pore, which can control the viscosity of the solution and then the ionic mobility.[44] As another important physical property of the solution in the pore, the EOF was investigated in axial and radial directions. The radial speed distribution in the center cross section of the pore is plotted in Figure 6d. At low positive voltages, such as 0.2 and 0.4 V, the EOF followed the forward direction which stemmed from the positive electrode to the negative electrode, and was weak due to the high viscosity of the solution in the pore. As the voltage increased, the magnitude of EOF first decreased nearly to 0 mm/s at 0.6 V, then switched direction and increased with the voltage. Because the solution viscosity was relatively high, the fluid flow caused by the directional movement of ions in electric double layers could not affect the flow velocity in the center of the pore in a significant manner. The reversed electroosmotic flow induced in the center of the pore moved from the negative electrode to the positive electrode, which brought the pore a solution of lower viscosity from the exit side (Figure S9b). As a result, the ionic mobility increased under high positive voltages which produced a much larger current as we found in the experiments. In Figure 6d, the velocity distribution of EOF in radial direction was similar to that found before under concentration gradients.[14] Based on the ionic concentration distributions, the reversed EOF was caused by the more $Cl^-$ ions in the center of the pore than K⁺ ions. From the Navier-Stokes equations, the net force of the fluid in the center region followed the moving direction of Cl⁻ ions. 2D EOF flow in the pore was plotted in Figure 7, which revealed that the direction of electroosmotic flow was dependent on the radial position at voltages higher than 0.6 V. Consequently, formation of vortices was expected, which would lead to a more efficient mixing of the two solutions.

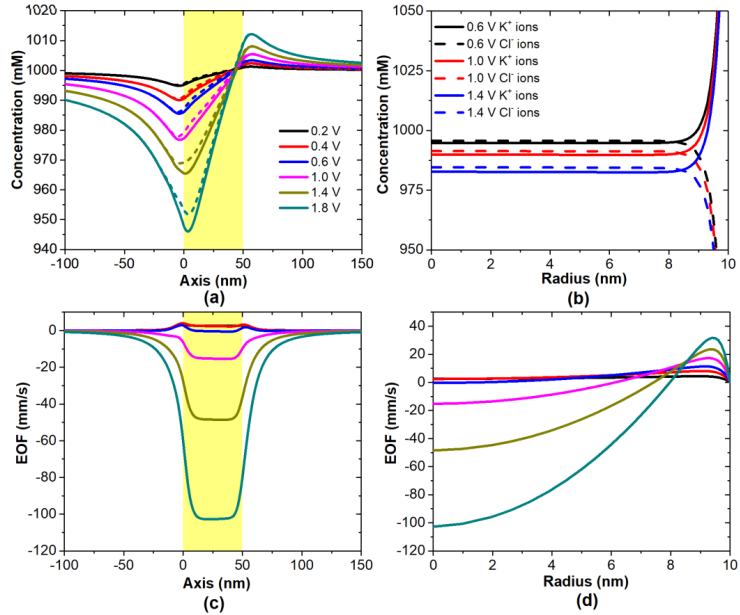

Figure 6. Simulation results from a nanopore with 50 nm in length and 20 nm in diameter. Concentration distributions of K⁺ (solid lines) and Cl⁻ (dashed lines) ions along the pore axis (a) and in the radial direction at the center cross section (b) of the pore. Electroosmotic flow along the axis of the pore (c) and in radial direction at the center cross section (d) of the pore. The viscosity gradient was set between 40% glycerol and aqueous solutions. Surface charge density of the pore walls was set as –0.01 C/m². The pore region is shown in yellow.

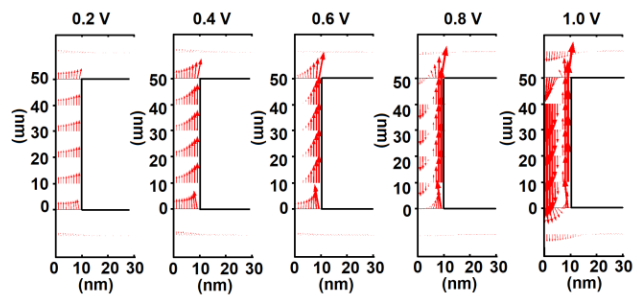

Figure 7. 2D distribution of electroosmotic flow in the pore with 20 nm in diameter and 50 nm in length at different voltages and under a viscosity gradient between 40% glycerol and aqueous solution of 1 M KCl. Surface charge density on the pore walls was set as –0.01 C/m$^2$.

In order to confirm the voltage-dependent direction of electroosmotic flow in short nanopores subjected to viscosity gradients, resistive-pulse detection[3, 51] using Au nanoparticles with 10 nm core and PEG$_{5000}$ modification had been conducted to show the direction of fluidic movement. The mean particle diameter was determined as 33.8 nm through dynamic light scattering method (Figure S11).[52] Due to the presence of PEG molecules on the particles surface, we assumed the particles were neutral. Experiments were conducted using small pores with ~20 nm in diameter due to the limitation of the current range as ±200 nA in the Axopatch 200B. Consequently, under symmetric aqueous condition, we did not observe particles' passages through the pore as typically done, but rather their approach to the pore opening, which also caused a small current decrease[53] as shown in Figure S12. This current drop was due to the change of resistance caused by the occupation of the particle in the access resistance regions.[54] Because the particles were only put in the cis chamber, the current drops were observed only for one voltage polarity consistent with the direction of electroosmotic flow. Please note that there may be few spikes under positive voltages due to the passed ultra-small particles. When solution in the trans chamber was replaced by 40% glycerol solution, at low positive voltages, similar current blockades were observed as those in aqueous solutions for one voltage polarity only (Figure S13a and 13b). With the voltage increasing to 1.5 V, resistive pulses appeared under both negative and positive voltages. In order to avoid the influence of the ultra-small particles passed through the pore, detection was also conducted with newly changed 40% glycerol solution on the trans side under reversed voltages. As shown in Figure S13d, resistive pulses also appeared.[53] Passage of particles for both voltage polarities could be explained by the radially inhomogeneous direction of electroosmotic flow shown in Figure 7. If a particle approached the pore close to the walls, it would be moved towards the pore opening in the direction of cations; if a particle approaches the pore closer to the pore center, it would experience the direction of reversed electroosmotic flow.

Finally, we wanted to identify parameters, which would allow us to tune the occurrence of the reversed electroosmotic flow, and rectification degree. Note that for the 50 nm long and 20 nm in diameter nanopore considered in Figure 3 the maximum rectification degrees obtained as ~1.9 and ~1.5 in the experiment and simulation were lower than the viscosity ratio of 3.98 and conductivity ratio of 2.87. Due to the dependence of the solution distribution in the pore on EOF which could be tuned with pore geometry and surface charge densities,[2] ionic transport was investigated with different pore geometries i.e. length and diameter, as well as different surface charge densities under viscosity gradient between 40% glycerol and aqueous solutions. As shown in Figure S14, under the same voltage, as we increased the nanopore length with 20 nm in diameter, ICR ratios increased as well in accordance with our prediction on smaller influence of the liquid-liquid miscible region on the pore conductance. For the pores with same length as 50 nm but different diameters, with the pore diameter increasing, ICR ratio became lower which was caused by the lower percentage of the pore resistance in larger pores due to the increased access resistance,[36] as well as the weaker viscosity gradient caused by the higher flux of EOF(Figure S15). Through changing the pore geometry, the ICR ratio still cannot reach the viscosity gradient within 3 V. Higher voltages and surface charge density had also been considered because the EOF depended on electric field strength and surface potential linearly.[2] Under the same electric field strength (applied voltage divided by pore length), we found that only for pores longer than 500 nm, the current rectification ratio reached the viscosity ratio under strong electric fields. While, super large voltages may not be feasible in experiments. Using a higher surface charge density on the pore surfaces, much stronger current rectification appeared across the pore within 3 V. For the case of a pore with –0.04 C/m$^2$, the ICR ratio could reach 3.5 approaching the viscosity gradient. As shown in Figure S16, the strong EOF filled most regions inside the pore with viscous solution, which also increased the viscosity outside the pore on the exit side. From the ICR ratio plots, the voltage corresponding to the peak value can roughly represent the turning point. With the increased pore length, enhanced surface charge density, and decreased pore diameter, the turning point located at larger voltages (Figure S14d), which was caused by the balance between electroosmotic flow and the reversed electroosmotic flow.

**Conclusions**

Nanopores with low aspect ratios present a good model system to investigate the liquid-liquid miscible region in confined spaces. Herein, with the method of current detection, ionic behaviors were explored under viscosity gradients between aqueous and glycerol solutions. With a viscosity gradient across thin nanopores, abnormal current rectification was found when the pore was filled with solutions of high viscosities under relative high electric field strengths. From current voltage curves, ionic current had a turning point, after which the current value increased much faster and the current rectification ratio decreased in a different trend from normal current rectifications. With the help from simulations, this abnormal current rectification was found to be caused by the reversed electroosmotic flow, which dragged the aqueous solution into the pore to lower the viscosity. The excessive coions due to their slower depletion than counterions along the pore can be responsible for the unexpected reversed electroosmotic flow in a negatively charged confined channel. Through enhancing the surface charge density of the pore, ionic current rectification ratio could reach the viscosity gradient across thin nanopores.

The experiments and modeling presented in this manuscript also revealed that enhanced mixing of solutions could occur in nanoconfined spaces if reversed EOF appeared. The induced mixing could find applications in designing nano-reactors and probing chemical reactions at the nanoscale. Meanwhile, the enhanced sensitivity of ICR of short nanopores to viscosity makes them a very attractive tool to dynamically probe local changes of the solution viscosity induced e.g. by products of chemical reaction or presence of molecules or particles.

## Acknowledgement

We acknowledge funding from the National Institutes of Health (HG009186). We thank M. Alibakhshi at Northeastern University for help in fabrication of SiN membrane chips and Prof. S. Sridhar at Northeastern University for use of the dynamic light scattering instrument.

## Supporting Information

Conductivity of the solutions in the experiments; Boundary conditions and parameters in the simulations; Viscosity obtained from simulations; Repeatability of abnormal ionic current rectification; Ionic and fluidic behaviors from simulation; Resistive-pulse detection of nanoparticles; Simulation results from pores with different sizes and surface charges; A simulation report.

TOC

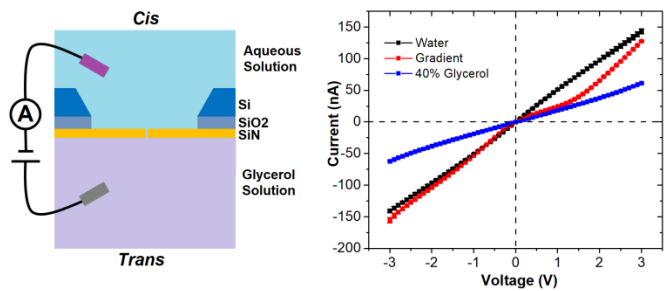